\documentclass[12pt]{article}

\usepackage{epsfig}
\usepackage{amssymb,amsmath}
\usepackage{graphicx, caption}
\usepackage{epstopdf}
\usepackage{setspace}

\newcommand{\bea}{\begin{eqnarray}}
\newcommand{\eea}{\end{eqnarray}}

 \makeatletter
\def\alt{\mathrel{\mathpalette\gl@align<}}
\def\agt{\mathrel{\mathpalette\gl@align>}}
\def\gl@align#1#2{\lower.6ex\vbox{\baselineskip\z@skip\lineskip\z@
\ialign{$\m@th#1\hfil##\hfil$\crcr#2\crcr\sim\crcr}}} \makeatother

\begin{document}
%

\begin{center}
\baselineskip 20pt {\Large\bf
$\mu$-Term Hybrid Inflation and Split Supersymmetry
}
\vspace{1cm}

{\large
Nobuchika Okada$^{a,}$\footnote{E-mail:okadan@ua.edu}
and Qaisar Shafi$^{b,}$\footnote{ E-mail:shafi@bartol.udel.edu}
} \vspace{.5cm}

{\baselineskip 20pt \it
$^b$Department of Physics and Astronomy,\\
University of Alabama, Tuscaloosa, AL 35487, USA \\
\vspace{2mm}
$^a$Bartol Research Institute, Department of Physics and Astronomy, \\
University of Delaware, Newark, DE 19716, USA
}
\vspace{.5cm}

\vspace{1.5cm} {\bf Abstract}
\end{center}

We consider $\mu$-term hybrid inflation which, in its minimal format with gravity mediated 
  supersymmetry breaking, leads to split supersymmetry. 
The MSSM $\mu$-term in this framework is larger than the gravitino mass $m_G$, 
  and successful inflation requires $m_G$ (and hence also $|\mu|$) $\gtrsim 5 \times 10^7$ GeV, 
  such that the gravitino decays before the LSP neutralino freezes out. 
Assuming universal scalar masses of the same order as $m_G$, 
  this leads to split supersymmetry. 
The LSP wino with mass $\simeq$ 2 TeV is a plausible dark matter candidate, 
  the gluino may be accessible at the LHC, and the MSSM parameter
  $\tan \beta \simeq 1.7$ in order to be compatible with the measured Higgs boson mass. 
The tensor-to-scalar ratio $r$, a canonical measure of gravity waves, can be as high as $0.001$. 

\thispagestyle{empty}

\newpage

\addtocounter{page}{-1}

\baselineskip 18pt

Minimal supersymmetric F-term hybrid inflation employs a renormalizable superpotential $W$ and
   a canonical K\"{a}hler potential $K$~\cite{DSS, EJC}.
The form of $W$ is determined by a U(1) R-symmetry, which contains the MSSM `matter' parity as a $Z_2$ subgroup.
In the supersymmetric limit an underlying gauge symmetry $G$ is spontaneously broken to a subgroup $H$.
With minimal $W$ and $K$, $G$ breaks to $H$ at the end of inflation.
The first calculations exploited quantum corrections induced by supersymmetry breaking to drive inflation,
  and the scalar spectral index in this case was estimated to be $n_s = 1 - 1/N \simeq 0.98$~\cite{DSS},
  where $N= 60$ denotes the number of e-foldings necessary to resolve the horizon and flatness problems of Big Bang Cosmology.

Subsequently, it was realized~\cite{RSW, buchm} that a class of linear soft supersymmetry breaking terms
  also should be included in the inflationary potential, and this allows improved agreement with scalar spectral index values
  of $n_s=0.96-0.97$ determined by the WMAP~\cite{WMAP9y} and Planck satellite experiments~\cite{Planck2015}.

The importance of this linear soft supersymmetry breaking term had previously been emphasized
 by Dvali, Lazarides and Shafi (DLS)~\cite{DLS} in the context of inflation and the MSSM $\mu$ problem.
The U(1) R-symmetry, following \cite{DLS}, prevents the appearance of the direct MSSM $\mu$ term.
The latter is generated after the inflaton field acquires a non-zero VEV as a consequence of supersymmetry breaking.
Assuming minimal $K$, the magnitude of $\mu$ is typically larger than the gravitino mass $m_G$~\cite{DLS}.

In this paper we explore the phenomenological implications that follow from implementing the DLS mechanism
   in the framework of minimal supersymmetric hybrid inflation.
We will find that a consistent inflationary scenario, taking into account reheating and the cosmological gravitino constraint,
   yields rather concrete predictions regarding supersymmetric dark matter and Large Hadron Collider (LHC) phenomenology.
In particular, the gravitino must be sufficiently heavy ($m_G \gtrsim 5 \times 10^7$ GeV),
   so that it decays before the freeze out of the lightest superpartner (LSP) neutralino dark matter.
A particularly compelling dark matter candidate turns out to be the wino with mass $\simeq 2$ TeV.
The soft scalar mass parameter $m_0$ is expected to be of the same order as $m_G$ or larger,
   which can reproduce a SM-like Higgs boson mass $\simeq 125$ GeV for suitable $\tan \beta$ values,
   where $\tan \beta$ is the ratio of the VEVs of the two MSSM Higgs doublets.
Depending on the underlying gauge symmetry $G$ associated with the inflationary scenario,
   the observed baryon asymmetry in the universe can be explained via leptogenesis \cite{Fuku-Yana, LS}.
Compelling examples of $G$ in which the DLS mechanism can be successfully merged with inflation include
   U(1)$_{B-L}$ and SU(2)$_L \times$SU(2)$_R \times$U(1)$_{B-L}$.
Other examples of $G$ are SU(5) and SU(4)$_c \times$SU(2)$_L \times$SU(2)$_R$~\cite{Pati-Salam}.
The latter, however, require extra care because of the presence of monopoles and we leave this for discussion elsewhere.

In the minimal supersymmetric model under discussion, inflation is associated with the spontaneous breaking of a gauge symmetry $G$.
A U(1) R-symmetry yields the following unique renormalizable superpotential:
\begin{equation}
     W = S ( \kappa \overline{\Phi}\Phi - \kappa M^2 + \lambda H_u H_d ).
\end{equation}
Here $\Phi$, $\bar{\Phi}$ denote a conjugate pair of chiral superfields whose VEVs induce the spontaneous breaking of $G$,
   with supersymmetry unbroken.
The dimensionless coefficients $\lambda$ and $\kappa$ can be made real by suitable field redefinitions.
The superpotential $W$ and the $G$-singlet superfield $S$ are assigned unit R-charges, while the remaining superfields have zero R-charges.

In the case of global supersymmetry, the VEV of the scalar component of the $G$-singlet superfield $S$ is zero.
However, taking into account supersymmetry breaking in supergravity, the $S$ field acquires a non-zero VEV proportional to the gravitino mass $m_G$.
Following \cite{DLS}, we have included the term $\lambda S H_u H_d$, which then induces the desired MSSM $\mu$ term.
It was shown in \cite{DLS} that to implement successful inflation, followed by the desired breaking of $G$ and
  the MSSM gauge symmetry unbroken, the parameter $\lambda > \kappa$.
This additional term in the superpotential will play an essential role in our analysis.

Assuming a canonical (minimal) K\"{a}hler potential, and taking into account radiative corrections and
   the linear soft supersymmetry breaking term, the inflationary potential is approximated by 
\bea
V(\phi) = m^4 \left( 1+ A \ln \left[\frac{\phi}{\phi_0} \right] \right) - 2 \sqrt{2} m_G m^2 \phi.
 \label{Vinf}
\eea
We have identified the inflaton field $\phi$ with the real part of $S$, the renormalization scale ($Q$) is set equal
   to the initial inflaton VEV $\phi_0$, and the imaginary part of $S$ is assumed to stay constant during inflation
[For a more complete discussion of this last point see Ref.~\cite{buchm}].
The negative sign of the linear term is essential to generate the correct value for the spectral index.
Note that in the absence of this linear term the scalar spectral index $n_s$ is predicted to lie close to $0.98$, as shown in \cite{DSS}.
Finally, $m = \sqrt{\kappa} M$ , and the coefficient $A \ll1$ is given by
\begin{equation}
A =  \frac{1}{4 \pi^2} \left( \lambda^2 + \frac{N_{\Phi}}{2} \kappa^2 \right).
\end{equation}
The coefficient $N_{\Phi}$ depends on the model under discussion.
For instance, $N_{\Phi} =1$ if $G =$U(1)$_{B-L}$, while it is 2 for $G=$SU(2)$_L \times$SU(2)$_R \times$U(1)$_{B-L}$.
Note that for inflaton field values close to $M$, the leading supergravity corrections are of order $(M/m_{P})^4$
  and therefore well suppressed, 
  and also the quadratic soft supersymmetry breaking term $m_\phi^2 \,\phi^2$
  associated with the inflaton can be ignored relative to the liner term in Eq.~(\ref{Vinf})~\cite{RSW, buchm}.

In the following discussion, where appropriate, we will set
\bea
V \simeq m^4, \; \;  \;
V^\prime = m^4 \; \frac{A}{\phi} \left( 1 - B \frac{\phi}{\phi_0} \right), \; \; \;
V^{\prime \prime}  = -m^4 \; \frac{A}{\phi^2},
\eea
where
\begin{equation} \label{beta}
 B = \frac{2\sqrt{2} \; m_G \; \phi_0}{A \; m^2}
\end{equation}
is a dimensionless parameter of order unity.
Note that $B = 0$ if the linear soft supersymmetry breaking term is ignored.
At the potential minimum, the inflaton VEV is given by $\langle \phi \rangle$ = $m_G / \kappa$,
   which yields the MSSM $\mu$-term: 
\begin{equation}
   \mu  = \gamma m_G,
\end{equation}
where $\gamma \equiv \lambda/\kappa > 1$.

The well-known slow roll parameters are given by
\bea
\epsilon = \frac{1}{2}  \left( \frac{V'}{V} \right)^2  \simeq \frac{1}{2}  \; \frac{A^2}{\phi^2} \left(1-B \frac{\phi}{\phi_0}\right)^2, \;\;\;
\eta = \frac{V''}{V} \simeq -\frac{A}{\phi^2}.
\label{slow_role}
\eea
For convenience we have used units where the reduced Planck mass $m_P = 2.43 \times 10^{18}$ GeV is set equal to unity.
For $A \ll 1$, $|\eta| \gg \epsilon$.
The end of inflation is determined by $|\eta|$ = 1, so that the inflaton value at the end of inflation
  is given by $\phi_e \simeq \sqrt{A}$.

\begin{figure}[ht]
  \begin{center}
   \includegraphics[width=10cm]{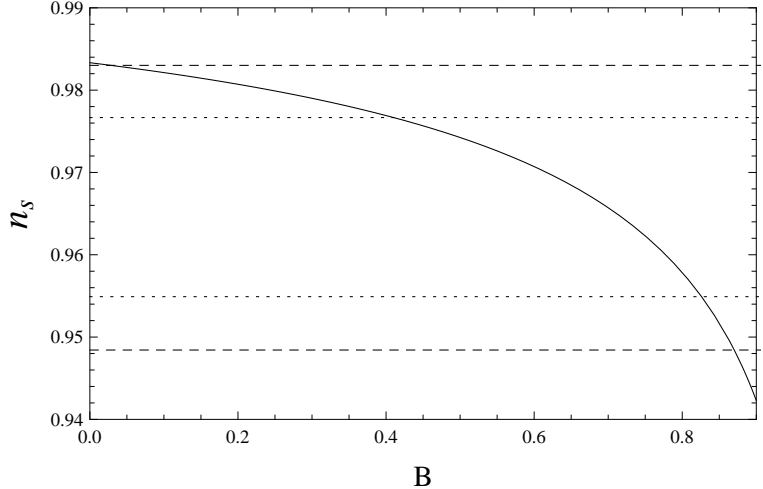}
   \end{center}
\caption{
Spectral index $n_s$ vs. $B$ as defined in Eq.~(\ref{beta}).
The region between the two dotted (dashed) lines corresponds to $1\sigma$ ($2 \sigma$) limit
  obtained by Planck 2015~\cite{Planck2015}.
}
 \label{fig:1}
\end{figure}

The number of e-foldings is estimated to be
\bea
N = \int_{\phi_e}^{\phi_0} \frac{V}{V'} \, d\phi
\simeq   \frac{\phi_0^2}{A} \int_{\phi_e/\phi_0}^1 \frac{x \, dx}{1-B x}
\simeq \frac{\phi_0^2}{A} \left( -\frac{1}{B} - \frac{\ln(1-B)}{B^2} \right)
\equiv \frac{\phi_0^2}{A} f(B),
\label{e-folds}
\eea
where we have used the approximation $\phi_e/\phi_0=0$ in the integral.
For the scalar spectral index, we obtain
\begin{equation}
n_s = 1 - 6\epsilon(\phi_0) + 2 \eta(\phi_0) \simeq  1 + 2 \eta(\phi_0) \simeq 1-\frac{2}{N} f(B).
\label{ns}
\end{equation}
In particular, for $N = 60$ and $B = 0$, $f (B) = 1/2$ and we obtain the well known result of Ref.~\cite{DSS}
  that $n_s= 1 - 1/N \simeq 0.983$.
To obtain the spectral index in the range of $0.955 \leq n_s \leq 0.977$
  which is indicated by $1\sigma$ region in the Planck 2015 results,
  the parameter $B$ lies in the range $0.83 \geq B \geq 0.41$, with $f (B)$ close to unity.
A plot of $n_s$ versus $B$ is shown in Fig.~\ref{fig:1}, along with 1$\sigma$ (dotted lines)
  and 2$\sigma$ (dashed lines) limits by the Planck 2015 results~\cite{Planck2015}.

The tensor-to-scalar ratio is given by
\bea
r=16 \epsilon(\phi_0)  \simeq 4 A (1-B)^2 (1-n_s),
\eea
where we have used Eqs.~(\ref{slow_role}) and (\ref{ns}).
With $0.83 \geq B \geq 0.41$ which yields the $n_s$ prediction within $1\sigma$ limit of the Planck 2015 results,
  we find
\bea
  r = (0.00014-0.00082) \times \left( \lambda^2 + \frac{N_{\Phi}}{2} \kappa^2 \right).
\label{tts}
\eea

To proceed further we employ the power spectrum whose magnitude has been estimated (Planck 2015) to be
\begin{equation}
  \Delta_{\mathcal{R}}^2 = \frac{1}{24 \pi^2} \frac{V(\phi_0)}{\epsilon(\phi_0)} = 2.20 \times 10^{-9}.
\end{equation}
\noindent
Utilizing $V \simeq m^4$, and Eqs.~(\ref{slow_role}) and (\ref{e-folds}), we find the following expression 
  for the gauge symmetry breaking scale $M$:
\begin{equation}
M \simeq 0.0032 \times  \sqrt{\tilde{\gamma}} \left(\frac{60}{N}\right)^{1/4} (1-B)^{1/2} f(B)^{1/4},
\end{equation}
where $\tilde{\gamma} =  \sqrt{\gamma^2 + N_{\Phi}/2}$.
Setting $N = 60$ and $0.83 \geq B \geq 0.41$, we estimate the symmetry breaking scale $M$:
\begin{equation}
M[{\rm GeV}] = (3.5 - 5.5) \times 10^{15} \sqrt{\tilde{\gamma}}.
\label{M_scale}
\end{equation}

The inflaton trajectory must be bounded along the $\Phi$, $\overline{\Phi}$ directions,
  which requires that $\phi_0 > M$. The ratio $\phi_0 / M$ is given by
\begin{equation}
\phi_0 / M \simeq 380 \times  \kappa \sqrt{\tilde{\gamma}} \left( \frac{N}{60} \right)^{3/4} (1-B)^{-1/2} f(B)^{-3/4}
\simeq (650 - 730) \times \kappa \sqrt{\tilde{\gamma}}  
\end{equation}
 for $0.83 \geq B \geq 0.41$, and requiring this ratio to be greater than unity imposes the following condition
on the dimensionless couplings $\kappa$ and $\lambda$:
\begin{equation}
\kappa \sqrt{\tilde{\gamma}} \; > (1.4-1.5) \times 10^{-3}.
\end{equation}

We next turn to inflaton decay into Higgsinos induced by the $\mu$-term in the superpotential (see paper 2 in Ref.~\cite{buchm}).
The decay width is estimated to be
\bea
\Gamma (\phi_0 \to  \tilde{H_u} \tilde{H_d}) = \frac{\lambda^2}{8 \pi} m_{\phi},
\eea
where $m_\phi = \sqrt{2} \kappa M$ is the inflaton mass.
We estimate the reheating temperature after inflation $T_{RH}$ 
  using $\Gamma = H =  T_{RH}^2\sqrt{\frac{\pi^2}{90} g_{\star}} $ with $g_{\star} = 228.75$ for the MSSM, 
  which gives
\bea
%
T_{RH} [{\rm GeV}] \simeq 1.5 \times 10^{16}   ( \kappa \sqrt{\tilde{\gamma}} )^{3/2} \frac{\gamma}{\sqrt{\tilde{\gamma}}}
\left( \frac{228.75}{g_{\star}} \right) \left( \frac{60}{N} \right)^{1/8} (1-B)^{1/4} f(B)^{1/8}.
\eea
The condition, $\phi_0/M >1$, leads to
\bea
  T_{RH} [{\rm GeV}] >  (5.0 - 7.5) \times 10^{11} \frac{\gamma}{\sqrt{\tilde{\gamma}}}
\eea
  for the $1 \sigma$ limit from Planck 2015.
Using $\tilde{\gamma} = \sqrt{\gamma^2 + N_{\Phi}/2}$ and $\gamma >1$,
  we obtain the lower bound $T_{RH} \gtrsim (5.0 - 7.5) \times 10^{11}$ GeV.

To discuss cosmology in the presence of gravitino, we first consider the case that the latter is 
  the LSP and hence a potential dark matter candidate.
In this case, the relic density of thermally produced gravitino is estimated as \cite{gravitino_abundance}
\bea
\Omega_{G}h^2 \simeq 0.21 \left( \frac{T_{RH}}{10^{11}\text{ GeV}} \right)
\left( \frac{1 \text{ TeV}}{m_G} \right) \left( \frac{M_3}{1 \text{ TeV}} \right)^2,
\eea
 where $M_3$ is the gluino mass.
In order to reproduce the observed dark matter density ($\Omega_{G} h^2 =0.11$) and
  satisfy the lower bound $T_{RH} \gtrsim (5.0 - 7.5) \times 10^{11}$ GeV, we find
\bea
 \frac{m_G}{1 \text{ TeV}}  \gtrsim 10 \left( \frac{M_3}{1 \text{ TeV}}  \right)^2 .
\eea
For the current lower bound $M_3 \gtrsim 1$ TeV from the search for supersymmetry at the LHC~\cite{LHC_SUSY},
  we obtain $m_G > M_3$, which contradicts our original assumption that the gravitino is the LSP.
Therefore, we exclude the possibility of LSP gravitino.

If the gravitino is unstable, we encounter the cosmological gravitino problem~\cite{gravitino_problem},
  which originates from the gravitino lifetime,
\bea \label{taug}
 \tau_G \simeq 10^{4} \; {\rm sec} \times \left(  \frac{1\; {\rm TeV}}{m_G}\right)^3.
\eea
For $m_G < 21.5$ TeV, $\tau_G >1$ sec, and the gravitino decays after Big Bang Nucleosynthesis (BBN).
In this case, the energetic particles created by the gravitino decay can destroy the light nuclei
 successfully synthesized during BBN.
To avoid this problem, the reheating temperature after inflation has an upper bound,
  $T_{RH} < 10^6-10^9$ GeV for $100$ GeV$\lesssim m_G \lesssim10$ TeV~\cite{Trh_BBN}.
The lower bound on reheating temperature in our scenario is not consistent with this upper bound.

For $m_G > 21.5$ TeV, the BBN bound on reheating temperature is not applicable
  since the gravitino decays before BBN.
However, in this case, we need to consider another cosmological constraint that the relic density of the LSP neutralino
  produced by gravitino decay should not exceed the observed dark matter relic density $\Omega_{{\tilde \chi}^0} h^2 = 0.11$.
Because of R-parity conservation, the LSP number density is the same as the number density of gravitino, and
  the upper bound on $\Omega_{{\tilde \chi}^0} h^2$ is expressed as $m_{{\tilde \chi}^0}  Y_G \leq 4\times 10^{-10}$~\cite{Trh_BBN},
  where $m_{{\tilde \chi}^0}$ is the LSP neutralino mass, and
\bea
  Y_G \simeq 10^{-11} \left( \frac{T_{RH}}{10^{10} \;{\rm GeV}}  \right)
\eea
is the yield of gravitino.
For the lower bound $T_{RH} \gtrsim (5.0 - 7.5) \times 10^{11}$ GeV, we find $m_{{\tilde \chi}^0}\lesssim 1$ GeV.
Although such a light neutralino is still consistent with laboratory constraints~\cite{vlLSP},
  its thermal relic abundance is found to be $\Omega_{{\tilde \chi}^0} h^2 \gg 0.1$
  for all the other superpartner masses $\gtrsim100$ GeV~\cite{HT} .

To avoid the constraint on the neutralino abundance from gravitino decay, 
   assume that the LSP neutralino is still in thermal equilibrium when the gravitino decays.
In this case, the LSP neutralino abundance is not related to the gravitino yield.
Using a typical value of the ratio $x_F\equiv m_{{\tilde \chi}^0}/T_F \simeq 20$,
  where $T_F$ is the freeze out temperature of the LSP neutralino, this occurs for the gravitino lifetime,
\bea
  \tau_G \lesssim 4 \times 10^{-10} \left( \frac{1 \text{ TeV}}{m_{{\tilde \chi}^0}} \right)^2.
\eea
Combining this with Eq.(\ref{taug}), we find
\bea
  m_G \gtrsim 4.6 \times 10^7 \text{ GeV} \left( \frac{m_{{\tilde \chi}^0}}{2 \text{ TeV}} \right)^{2/3}.
\label{LB_mG}
\eea
Therefore, our cosmological scenario favors a gravitino mass at an intermediate scale.

In our $\mu$-term generation mechanism, $\mu = \gamma m_G$ with $\gamma \gtrsim 1$.
To implement the electroweak symmetry breaking, the soft supersymmetry breaking mass for the MSSM Higgs doublets
  should satisfy $|m_0^2| \gtrsim \mu^2$.
Thus, the soft scalar mass parameter $m_0$ also lies at an intermediate scale.
On the other hand, the mass scale of the dark matter neutralino is of order 100 GeV$-$TeV
  to reproduce the observed relic abundance.
Therefore, the so-called split SUSY scenario~\cite{splitSUSY}  is compatible to our scheme, 
  and we expect a hierarchy between the soft supersymmetry breaking scalar masses and the gaugino masses
  [A special structure in supergravity is crucial to realize this mass splitting~\cite{LO}].
In split SUSY with a large $\mu$, an LSP wino with mass around 2 TeV is the simplest dark matter candidate~\cite{splitSUSy-pheno}.

\begin{figure}[h!]
  \begin{center}
   \includegraphics[width=10cm]{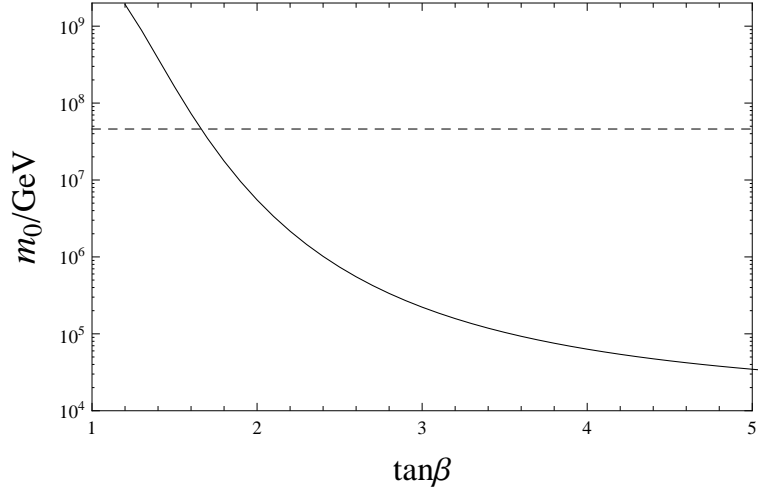}
   \end{center}
\caption{
Soft scalar mass ($m_0$) as a function of $\tan \beta$.
Along the solid line, the Higgs boson mass $m_h=125.09$ GeV is realized
 for the top quark mass input $M_t=173.34$ GeV.
The horizontal dashed line denotes $m_0=4.6 \times 10^7$ GeV,
 corresponding to the lower bound on gravitino mass in Eq.~(\ref{LB_mG})
 with $m_{{\tilde \chi}^0}=2$ TeV.
}
  \label{fig:2}
\end{figure}

Let us fix the scalar mass scale in split SUSY so as to realize the observed Higgs boson mass.
With the decoupling of heavy scalar fields at a typical mass scale $m_0$, split SUSY provides
  a boundary condition for the SM Higgs quartic coupling at $Q=m_0$~\cite{splitSUSy-pheno},
\bea
  \lambda_h(m_0) = \frac{1}{4} \left(  g(m_0)^2 + g^\prime(m_0)^2 \right) \cos^2 2 \beta.
\label{BCforlambda}
\eea
We employ the SM renormalization group equations at two-loop level
   with the boundary conditions at top quark pole mass ($M_t$) presented in \cite{SM_RGE}.
As inputs, we use the central value of the combination of Tevatron and LHC measurements of top quark mass
   $M_t = 173.34$ GeV~\cite{Mt}, and the central value of the combined analysis for Higgs boson mass measurements
   by the ATLAS and the CMS~\cite{mh_combined}, $ m_h = 125.09$  GeV,
   along with $m_W=80.384$ GeV and $\alpha_s=0.1184$.
Using numerical solutions to the renormalization group equations of
   the SM Higgs quartic coupling and the electroweak gauge couplings,
   we find $m_0$ which satisfies the boundary condition of Eq.~(\ref{BCforlambda}) for a fixed $\tan \beta$.

Our result is depicted in Fig.~\ref{fig:2}, where the Higgs boson mass $m_h=125.09$ GeV is reproduced
    along the solid line.
For $m_0=4.6 \times 10^7$ GeV, corresponding to the lower bound on gravitino mass
  in Eq.~(\ref{LB_mG}) with $m_{{\tilde \chi}^0}=2$ TeV, $m_h = 125.09$  GeV is reproduced for $\tan \beta \simeq 1.7$.

Since the gauge group associated with inflation is an extension of the SM one, it may be natural to set 
  the symmetry breaking scale $M$ equal to the gauge coupling unification scale, $M \simeq 2 \times 10^{16}$ GeV,
  which is achieved with ${\tilde\gamma}=13-32$ in Eq.~(\ref{M_scale}).
In this case, we may choose $\lambda \simeq 1$ and $\kappa \simeq 0.1$, for which $\phi_0/M \gg 1$.
For $\lambda=1$, the tensor-to-scalar ratio $r=(1.4-8.2) \times 10^{-4}$
  for $0.83 \geq B \geq 0.41$ (see Eq.~(\ref{tts})).
With somewhat larger values of $\lambda$, $r$ can be as high as $10^{-3}$. 

In summary, the coupling of the MSSM Higgs doublets to the inflaton field induces the $\mu$-term
  and also leads to important predictions concerning the gravitino, neutralino dark matter, 
  Higgs boson mass and low energy phenomenology.
A consistent inflationary scenario requires a fairly large gravitino mass ($m_G \gtrsim 5 \times 10^7$ GeV),
  and in the simplest scheme a 2 TeV wino is a compelling thermal dark matter candidate.
The soft supersymmetry breaking scalar masses are of order $10^7$ GeV, with $\tan \beta \simeq 1.7$,
  in order to generate the desired 125 GeV mass for the SM-like Higgs boson.
The Higgsinos are rather heavy because $\mu$ is predicted to be larger than the gravitino mass.

\section*{Acknowledgments}
N.O. would like to thank the Particle Theory Group of the University of Delaware for hospitality during his visit. Q.S. thanks Nefer Senoguz, George Lazarides, Costas Pallis, and Dylan Spence for discussions.
This work is supported in part by the DOE Grant No.~DE-SC0013680 (N.O.) and No.~DE-SC0013880 (Q.S.).


\end{document}